\documentclass[prl,twocolumn,showpacs,aps]{revtex4}
\usepackage{epsfig}
\newcommand{\abs}[1]{\left|#1\right|}
\newcommand{\rmi}{{\rm i}}

\newcommand{\op}[1]{\hat{#1}}
\newcommand{\sign}[1]{\rm{sign}(#1)}
\newcommand{\mean}[1]{\left\langle#1\right\rangle}

\begin{document}

\title{Atom optics kicked rotor: experimental evidence for a pendulum description of
the quantum resonance}
\author{S. Wayper, M. Sadgrove, W. Simpson, and M.D. Hoogerland }
\affiliation{Department of Physics, University of Auckland, Private Bag 92019,
Auckland, New Zealand}

\begin{abstract}
We present measurements of the mean energy for an atom optics kicked rotor ensemble
close to quantum resonance. Oscillations in the mean energy in this regime are
are shown to be in agreement with a quasi--classical pendulum approximation.
The period of the oscillations is shown to scale with a single variable, which
depends on the number of kicks.
\end{abstract}

\maketitle
\newcommand{\kbar}{\mathchar'26\mkern-9muk}

The `kicked rotor' system has provided an extremely useful model for the
study of the correspondence between classical and quantum dynamics for almost
three decades. This system behaves as a free rotor except during periodically
applied momentum kicks, when it experiences a position dependent potential
giving
rise to sharp changes in momentum. Classically, such a system exhibits essentially
stochastic (i.e. \emph{chaotic}) diffusion if the potential strength is
large enough 
\cite{Rechester1980,Lichtenberg1992}.
The quantised system, however, avoids this signature of chaos in two
principal ways:
Firstly, the celebrated phenomenon of \emph{quantum dynamical localisation} may
occur
in which quantum interference between components of the wavefunction for individual 
kicked atoms leads to saturation in the energy growth of the ensemble \cite{stddylocrefs,Moore1995}.
Secondly, for certain kicking frequencies, quadratic energy growth may occur for atoms 
with the correct
initial momentum in a phenomenon known as \textit{quantum resonance} (QR)
\cite{Izrailev1979,Oskay2000}.
Both of these quantum features are plainly at odds with a
classical interpretation, 
the first because
chaotic diffusion is inhibited, and the second because a completely chaotic
system cannot be driven
on resonance. 

In this paper we are concerned with the structure and dynamics of the enhanced
energy peaks seen near the quantum resonances of the quantum kicked rotor.
For the first time, we present firm experimental evidence of the validity of
a quasi-classical pendulum description of the system. In our experiment,
we
employ an ensemble of laser-cooled atoms which interact with a 
standing wave of laser light, which is detuned from resonance and pulsed
periodically.  
The AC Stark shift induced by the laser field forms a
potential, which depends sinusoidally on the position in the standing
wave, leading to dynamics which are formally identical to the kicked rotor.
This realisation of the model kicked rotor system is known as the \emph{atom optics kicked rotor} 
(AOKR)\cite{Moore1995}. For certain pulse periods, in this system, enhanced energy peaks arise,
as reported in Refs. \cite{dArcy2001}. 
Recently, a quasi--classical approximation to the quantum dynamics, which is
valid near exact quantum resonance
\cite{Wimberger2003} has been developed. This theory, known as
$\epsilon$--classical dynamics, accurately predicts the
behaviour of kicked atoms near QR \cite{Wimberger2005} utilising a pendulum
approximation, which is closely related to, although not equivalent to, the
resonance in the classical
limit of the quantum kicked rotor \cite{Sadgrove2005}. In this paper we measure the oscillations
to either side of the main quantum resonance peak that are predicted by the
$\epsilon$--classical pendulum approximation. The oscillations are found to move
towards the resonance peak with the time dependence predicted by the pendulum scaling of
the motion, providing the strongest confirmation yet of the validity of $\epsilon$--classical
dynamics in the neighbourhood of the the quantum resonance. The excellent
agreement of such details between the $\epsilon$--classical model and a
``quantum'' feature of the experiment signifies a new level of understanding of
quantum dynamics of this simple system. 

In a quantum-mechanical picture, the Hamiltonian for the AOKR kicked with period
$T$ by an optical standing wave
having a wavenumber $k_l$ is given by
\begin{equation}
{\cal H }(t') =\frac{\kbar\hat{p}^2}{2} + k\cos(\hat{x})\sum_{n=0}^{N}
 \delta (t'-n)\;,
\label{eq:ham}
\end{equation}
where $\hat{x}$ is the atomic position operator scaled by $2 k_l$,
$\hat{p}$ is the momentum operator in units of $2\hbar k_l$,
$k$ is the kicking strength, the scaled time is $t'=t/T$ and $n$ 
is an integer which counts the number of kicks. The quantity $\kbar$ 
may be viewed as a scaled Planck's
constant and is
defined by the commutator relation $[\hat{x}, \hat{\rho}] = i\kbar$, 
where $\kbar=8\omega_rT$ ($\omega_r$ is the frequency associated with the energy
 change after a single
photon recoil for Rubidum for a photon with $\lambda=780$nm).
We may write the momentum as $\hat{p} = (\hat{n}+\beta)$, 
where $\hat{n}$ the integer
momentum operator and $\beta$ the quasimomentum (or noninteger part of the momentum) which
is conserved during kicking.
The $\delta$--kick approximation is quite good in the experiments described
in this paper as the pulse width was $320$ns compared with a period of $\approx 30\mu$s.

We now present
a brief summary of the $\epsilon$--classical theory of the quantum resonance
peaks as it applies to our experiments, following the notation of Refs.
\cite{Wimberger2003}.
The Floquet operator for our system is
\cite{Izrailev1979,Bharucha1999,Wimberger2003}
\begin{equation}
\hat{U} = \exp({\rm i}k\cos\hat{\hat{x}})\exp(-{\rm i}\kbar \hat{p}^2/2).
\label{eq:evol}
\end{equation}
In Ref. \cite{Wimberger2003} it was shown that the evolution
operator (\ref{eq:evol}) could be written in an equivalent form:
\begin{equation}
\label{eq:evol_beta}
\hat{U}_\beta = \exp\left(-\rmi\frac{\tilde{k}}{\abs{\epsilon}}\cos(\hat{\theta})\right)
\exp\left(-\frac{\rmi}{\abs{\epsilon}}\hat{H}_\beta\right),
\end{equation}
where $\epsilon = \kbar - 2\pi\ell$ for integer $\ell$, $\tilde{k}=\abs{\epsilon}k$ and
 $\hat{H}_\beta$
is a function of $\beta$ and the new momentum operator $\hat{I}= \abs{\epsilon}\op{n}$ given by
$\op{H}_\beta = \frac{1}{2}\sign{\epsilon}\op{I}^2 
+ \op{I}(\pi\ell + \kbar\beta)$. We see that in Eq. \ref{eq:evol_beta}, the quantity
$\abs{\epsilon}$ plays the part of Planck's constant. However, $\abs{\epsilon}\rightarrow 0$
is a fictitious classical limit which corresponds to approaching exact quantum resonance.
For $\kbar$ close to $2\pi\ell$, therefore, the quantum dynamics is approximated well
by the \emph{classical} map \cite{Wimberger2003}
\begin{eqnarray}
\label{eq:ecsm}
J_{t+1} & = & J_t + \abs{\epsilon}k\sin(\vartheta_{t}),\label{eq:ecsmp}\nonumber \\
\vartheta_{t+1} & = & \vartheta_t + J_{t+1},\label{eq:ecsmx}
\end{eqnarray}
where  $J = \pm I + \pi\ell + \tau\beta$ and $\vartheta = \theta + \pi(1-\sign{\epsilon})/2$
with the mean energy given by $\mean{E}=\abs{\epsilon}^{-2}\mean{(J-J_0)^2/2}$.

\begin{figure}
\includegraphics[height=7cm]{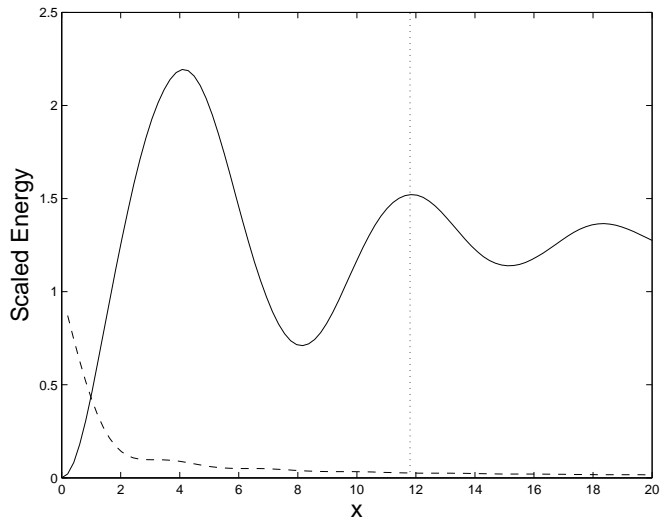}
\caption{\label{fig:Gfunc} The function $G(x)$ as given by Eq. \ref{eq:G}.
Note the secondary maximum
at $x\approx11.8$ (marked by the dotted line) which is responsible for the 
structure to either side of quantum resonance. For larger $x$ the 
function saturates to a value $\alpha \approx 1.3$. The dashed line shows $1-\Phi_0(x)$.}
\end{figure}

In the limit as $\abs{\epsilon}\rightarrow 0$, Wimberger \emph{et al.} showed
that
the mean energy of a kicked atomic ensemble is $\mean{E_t}=k^2t/4$.
For $\abs{\epsilon}>0$, a pendulum approximation to the dynamics of the
map \ref{eq:ecsm} is appropriate \cite{Lichtenberg1992}, with the
pendulum Hamiltonian $H_{\rm pend.} = J^2/2 + \abs{\epsilon}k\cos(\vartheta)$. The
pendulum motion then has a characteristic resonance period 
$t_{\rm res}=1/\sqrt{k\abs{\epsilon}}$.
The off--resonant mean energy, when scaled by the peak energy $\mean{E_{t,\epsilon}}=k^2t/4$
 may be written in terms of $x=t/t_{\rm res}$ as \cite{WimbergerPhD}
\begin{equation}
\label{eq:E_offres}
\frac{\mean{E_{t,\epsilon}}}{\mean{E_{t,\epsilon=0}}} \approx \left(1 - \Phi_0(x)\right)
 +\frac{4}{\pi x}G(x),
\end{equation}
where the term $G(x)$ is the energy due to the pendulum motion.
The term $1-\Phi_0(x)$ is the energy due to the remaining phase space
which is not affected by the pendulum resonance. It decays rapidly for
$t/t_{\rm res}\lesssim 1$ (see Fig. \ref{fig:Gfunc})and thus for $x\gg1$ we may write
\begin{equation}
\label{eq:Epend}
\frac{\mean{E_{t,\epsilon}}}{\mean{E_{t,\epsilon=0}}}\approx \frac{4}{\pi x}G(x).
\end{equation}

The function $G(x)$ is given explicitly by
\begin{equation}
\label{eq:G}
G(x) = \frac{1}{8\pi}\int_0^{2\pi}{\rm d}\theta\int_{-2}^{2}{\rm d}J'_0 J'(x,\theta_0,J'_0)^2,
\end{equation}
where $J'$ is the general solution for the pendulum momentum 
in terms of the Jacobi elliptic functions (see, for example, Ref. \cite{Reichl1992})).
The function $G(x=t/t_{\rm res})$ is shown in Fig. \ref{fig:Gfunc}. 
We see that it consists of decaying oscillations, which appear at constant values
of $x$. Whilst the first local maximum of $G$ is not visible in scans of the
quantum resonance peak, due to the predominance of the term $1-\Phi_0(x)$,
the second maximum at constant $x=x_0\approx 11.8$ should be visible as
oscillations to either side of the principle quantum resonance peak.
Furthermore, since these
oscillations have a constant $x$ position, their position relative to the
quantum resonance should change as a function of time according to the equation
\begin{equation}
\label{eq:hornpos}
\abs{\epsilon} = \frac{x_0}{t^2k}.
\end{equation}

To confirm the validity of this classical pendulum approximation near quantum
resonance, we perform a kicked rotor experiment using ultracold
rubidium 85 atoms from a standard magneto-optical
trap~\cite{MOT}, which is loaded from a background vapour. After an
accumulation phase, the quadrupole magnetic field is turned off and the trap
lasers are switched to a larger detuning for a 3~ms cooling phase, after which
the trap lasers are extinguished. The temperature of the atom cloud at that
moment is $\sim~10~\mu K$. The atoms are then subjected to a periodic sequence
of kicks by the (linearly polarised) standing wave laser field. The atomic
ensemble is subsequently allowed to expand for 15~ms, after which the ensemble
fluorescence is imaged on a CCD camera (Apogee AP47p) by flashing on the
molasses lasers for 5~ms. During the imaging period, the atoms experience an `optical
molasses'  and hence do not move significantly. We obtain the positional
variances of the atomic
cloud, in both the direction of the kick laser (`kicked') and orthogonal to that
(`non-kicked'), numerically from the image and convert these to average velocity
squared, and from that to kinetic energy. We take the energy in the non-kicked
direction as the initial energy. The experiment is then repeated for different
kick periods or kick numbers. The average initial energy is determined from an
experimental run over a range of parameters, and is used to rescale the energy
ratio to an energy.
\begin{figure}
\includegraphics[width=7cm]{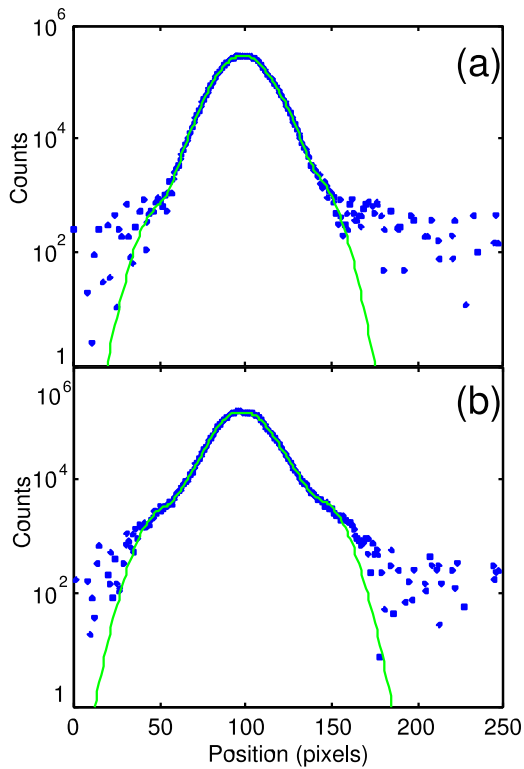}
\caption{Momentum distributions after 14 kicks, for kick periods of (a) 30.5~$\mu$s
(b) and 32.5~$\mu$s, corresponding to $\kbar=5.9$ and $\kbar=6.3$ respectively. 
The experimental data is shown as discrete (points). The solid lines and the result
of $\epsilon$-classical simulations. These results are for a scaled kick strength
$k=??$.\label{fig:pdist}}
\end{figure}

The kicking laser beam was obtained from a Toptica DLX110 laser
system, which was locked to the $D_2$ transition in $^{87}$Rb, obtaining a laser
detuning of 1.3~GHz from the $F=3\rightarrow F=4$ transition in $^{85}$Rb. The
beam was passed trough an AOM, which was controlled by a home-built programmable
pulse generator, allowing us to pulse the laser with adjustable pulse number,
period and amplitude. The pulse duration for most experiments was set to 320~ns.
The rise and fall times of the laser pulse were less than 50~ns.
To spatially filter the laser beam, it was then passed through a polarisation
preserving single mode fibre. After collimation, the radius $(1/e^2)$ of the
gaussian laser beam was 3.0~mm. The laser beam was passed through a polarising
beam splitter to ensure linear polarisation. The large detuning and linear laser
polarisation yield equal kick strengths for all magnetic sublevels of the $F=3$
ground state. The maximum kick laser power was 100 mW, which for a 320~ns pulse
and a 1.3~GHz laser detuning yields a scaled kick strength $k=5.2$. Larger
kick strengths could be obtained by lengthening the kick pulse. The detuning of
the kick laser yielded a negligible spontaneous emission rate for all laser
powers used. This was verified by observing the cloud energies for a range of
kick numbers, up to 80 kicks, which showed negligible energy growth
after a certain number of kicks. The initial radius of the atomic cloud was
0.3~mm ($1\sigma$), significantly smaller than the size of the kicking laser beam.

\begin{figure}
\includegraphics[width=6cm]{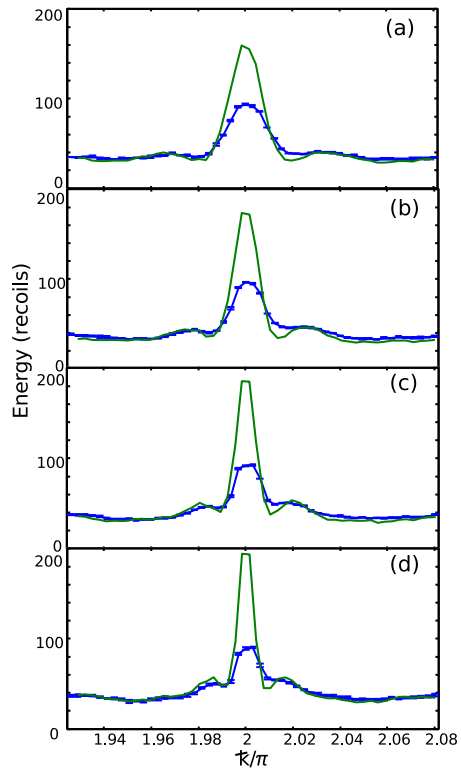}
\caption{Measured mean energies (discrete points) as a function of the kick period for
(a) 12 kicks, (b) 14 kicks, (c) 16 kicks and (d) 18 kicks. Solid lines correspond to
$\epsilon$-classical simulations. The scaled kick strength $k=4.2$. \label{fig:exp}}
\end{figure}

In figure~\ref{fig:pdist} the momentum distribution of the atoms is displayed
after 14 kicks, for a kicking period (a) close to to the first quantum resonance
($\kbar=2\pi$) and (b) exactly on resonance. For the on resonant case, a pedestal
is seen in the wings of the central peak corresponding to a small population
of resonant atoms. Experimentally, it is difficult to resolve the energy due
to these atoms but their relatively large momenta means that they contribute
greatly to the total mean energy. This lack of resolution (also noted by other groups
\cite{Wimberger2005})
causes the experimental values for the energy  on the quantum resonance to be
smaller than the theoretical values (see Fig. \ref{fig:exp}). For off resonant
kicking periods the pedestal is not present, and the experimental and simulated 
values of the mean energy show good agreement.

In figure~\ref{fig:exp}, the measured mean energy is displayed as a function of the
kicking period around the first primary quantum resonance, for different numbers
of kicks. It is interesting that an asymmetry in the peaks is found
 (also observed in \cite{Wimberger2005}).
This feature is most likely due to a small amount of spontaneous emission occuring with
each pulse. In fact, quantum simulations including the effect of spontaneous emission reproduce
this effect. The assymetry is, therefore, a purely experimental effect which is not allowed
for by the $\epsilon$--classical theory, but which does not destroy the basic structure
we seek to measure.

\begin{figure}
\includegraphics[width=9cm]{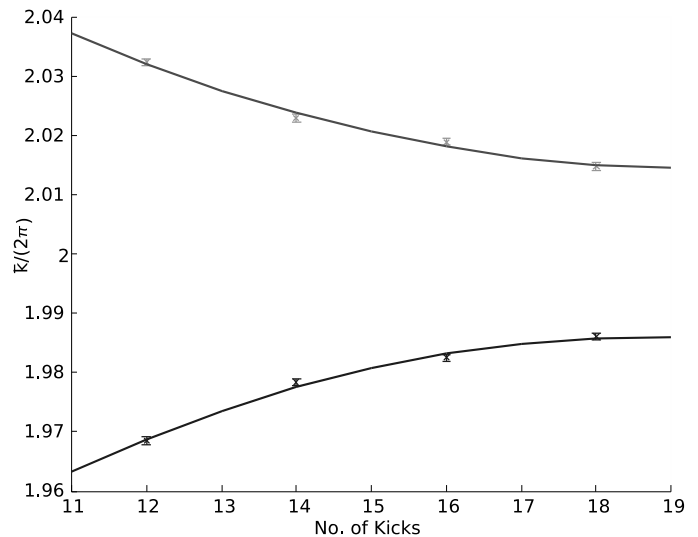}
\caption{\label{fig:peakmove} Peak position of the near resonant peaks plotted
against
kick number. The solid curve is that given by Eq. \ref{eq:hornpos} for
$k=4.1$. We see
excellent agreement between the predictions of the $\epsilon$--classical scaling
function
and the observed movement of the peaks as $N$ increases.}
\end{figure}

The experimental measurements are compared with epsilon classical simulations.
We find good agreement except near exact resonance where the mean energy 
calculation is affected by a lower signal--to--noise ratio for the small population
of resonant atoms. However the structure
of interest in these experiments lies away from the central resonance peak.
In particular, we draw attention to the small side peaks, which we are
resolved accurately here for the first time. These secondary maxima are expected to
exist if the pendulum scaling law (\ref{eq:Epend}) is correct. 
Furthermore, as shown in Figure \ref{fig:peakmove}, the side peaks move towards
the quantum resonance
as the number of kicks increases exactly as predicted by Eq. \ref{eq:hornpos} (choosing
$x_0=11.2$ to account for the fact that G is divided by x in the energy scaling).
We would like to emphasise the very accurate nature of the measurements performed here.
Fig. \ref{fig:peakmove} demonstrates that there is quantitative agreement between
the position of the side peaks as a function of kick number $t$ and Eq.
\ref{eq:hornpos}, 
providing strong
evidence that a classical pendulum scaling of the dynamics is valid in this regime.
These results show that the function $G(x)$ is an excellent predictor of the dynamical evolution
of the atom ensemble near quantum resonance

This confirmation of Eq. \ref{eq:hornpos} is also a direct way of demonstrating that the
resonance peak becomes narrower in a sub--Fourier manner, that is, at a rate faster than $1/t$.
This feature was also demonstrated in \cite{Wimberger2005} in \cite{Lille}
for other types of quantum resonances.

In conclusion, we have made the first careful measurements of the oscillations 
near quantum resonance, which are predicted by a classical pendulum scaling
law of the near resonant dynamics.
The structure itself is well resolved in our experiments. Furthermore, the
motion of the oscillations with kick
number has been observed to agree very well with the predictions of the
$\epsilon$--classical
theory.  These measurements set a new level of accuracy in the measurement of the rather 
subtle near resonant effects predicted by this new semi--classical approach to the dynamics
at quantum resonance.

MS acknowledges the support of a Top Achiever Doctoral scholarship 03131 and 
thanks Rainer Leonhardt for his assistance. The authors thank Sandro Wimberger
for illuminating discussions on theoretical matters.

\end{document}